\documentclass[reprint,superscriptaddress,amsmath,amssymb,prb,floatfix]{revtex4-2}

\usepackage[T1]{fontenc}
\usepackage[]{graphicx}
\usepackage{tabularx}
\usepackage[usenames,dvipsnames]{color}
\usepackage{soul}
\usepackage{bm}
\usepackage{amsmath}
\usepackage{lineno}
\usepackage{gensymb}

\usepackage{amsfonts}
\usepackage{mathrsfs}
\usepackage{graphicx}% Include figure files
\usepackage{dcolumn}% Align table columns on decimal point
\usepackage{bm}% bold math
\usepackage{color}

\usepackage[colorlinks,bookmarks=false,citecolor=blue,linkcolor=red,urlcolor=blue]{hyperref}
\usepackage{multirow}
\usepackage{physics}
\usepackage{siunitx}
\DeclareSIUnit{\barpressure}{bar}
\DeclareSIUnit\angstrom{\protect \text{Å}}

\begin{document}
	%TC:ignore	
	\title{Uncovering surface states of the Dirac semimetal BaMg\textsubscript{2}Bi\textsubscript{2}}
	% Force line breaks with \\
	\author{A. De Vita**}
    \affiliation{Institut für Physik und Astronomie, Technische Universität Berlin, Straße des 17 Juni 135, 10623 Berlin, Germany}
	\affiliation{Fritz Haber Institut der Max Planck Gesellshaft, Faradayweg 4--6, 14195 Berlin, Germany\looseness=-1}%

      \author{J. Bakkelund**}
	\affiliation{Department of Engineering Sciences, University of Agder, NO-4879 Grimstad, Norway}%

 \author{H. Świątek**}
	\affiliation{Faculty of Applied Physics and Mathematics, Advanced Materials Centre, Gdansk University of Technology, Narutowicza 11/12, 80-233, Gdansk, Poland\looseness=-1}%

  \author{M. J. Winiarski}
	\affiliation{Faculty of Applied Physics and Mathematics, Advanced Materials Centre, Gdansk University of Technology, Narutowicza 11/12, 80-233, Gdansk, Poland\looseness=-1}%

    \author{S. Malick}
	\affiliation{Faculty of Applied Physics and Mathematics, Advanced Materials Centre, Gdansk University of Technology, Narutowicza 11/12, 80-233, Gdansk, Poland\looseness=-1}%

    \author{C. V. B. Nielsen}
    \affiliation{Department of Physics and Astronomy, Interdisciplinary Nanoscience Center, Aarhus University, 8000 Aarhus C, Denmark}

    \author{F. Bertran}
    \affiliation{Synchrotron SOLEIL, L'Orme des Merisiers, Départementale 128, 91190 Saint-Aubin, France\looseness=-1}%

    \author{A. J. H. Jones}
    \affiliation{Department of Physics and Astronomy, Interdisciplinary Nanoscience Center, Aarhus University, 8000 Aarhus C, Denmark}

    \author{P. Majchrzak}
    \affiliation{Department of Applied Physics, Stanford University, Stanford, CA 94305-4090, USA}

    \author{F. Miletto Granozio}
    \affiliation{CNR-SPIN, c/o Complesso di Monte S. Angelo, IT-80126 Napoli, Italy}

    \author{J. A. Miwa}
    \affiliation{Department of Physics and Astronomy, Interdisciplinary Nanoscience Center, Aarhus University, 8000 Aarhus C, Denmark}

    \author{R. Ernstorfer}
    \affiliation{Institut für Physik und Astronomie, Technische Universität Berlin, Straße des 17 Juni 135, 10623 Berlin, Germany}
	\affiliation{Fritz Haber Institut der Max Planck Gesellshaft, Faradayweg 4--6, 14195 Berlin, Germany\looseness=-1}%

    \author{T. Pincelli}
    \affiliation{Institut für Physik und Astronomie, Technische Universität Berlin, Straße des 17 Juni 135, 10623 Berlin, Germany}
	\affiliation{Fritz Haber Institut der Max Planck Gesellshaft, Faradayweg 4--6, 14195 Berlin, Germany\looseness=-1}%
    
    \author{T. Klimczuk}
    \altaffiliation[Corresponding author: ]{tomasz.klimczuk@pg.edu.pl}%
	\affiliation{Faculty of Applied Physics and Mathematics, Advanced Materials Centre, Gdansk University of Technology, Narutowicza 11/12, 80-233, Gdansk, Poland\looseness=-1}%

    	\author{C. Bigi}
     \altaffiliation[Corresponding author: ]{chiara.bigi@synchrotron-soleil.fr}
	\affiliation{Synchrotron SOLEIL, L'Orme des Merisiers, Départementale 128, 91190 Saint-Aubin, France\looseness=-1}%

	\author{F. Mazzola}
	\altaffiliation[Corresponding author: ]{federico.mazzola@spin.cnr.it}%
    \affiliation{CNR-SPIN, c/o Complesso di Monte S. Angelo, IT-80126 Napoli, Italy}
%	\date{\today}% It is always \today, today,
	
	\begin{abstract}
		BaMg$_2$Bi$_2$ is a Dirac semimetal characterized by a simple Dirac cone crossing the Fermi level at the center of the Brillouin zone, protected by C$_3$ rotational symmetry. Together with its Sr-based analogue SrMg$_2$Bi$_2$, it has been proposed as a promising candidate for a chemically driven topological switch: while SrMg$_2$Bi$_2$ is an insulator, BaMg$_2$Bi$_2$ exhibits non-trivial topological features. A detailed understanding of its electronic structure is essential to elucidate its electronic and transport properties. Previous photoemission studies confirmed the Dirac nature of BaMg$_2$Bi$_2$, but were limited to high photon energies, which hindered direct comparison with density functional theory calculations (DFT), due to reduced resolution and higher-frequency matrix-element modulation in that regime. In this work, we combine high-resolution angle-resolved photoemission spectroscopy (ARPES) and DFT calculations to get full insight on the valence band states, providing a comprehensive picture of the low-energy electronic structure. Our measurements reveal the presence of previously unobserved surface states. We found that they are topologically trivial, but they unlock a more comprehensive understanding of the material's behavior, reconciling previous discrepancies between experiment and theory.
	\end{abstract}

\maketitle

\noindent ** These authors contributed equally.
      
      \begin{figure}
		\includegraphics[width=0.7\linewidth]{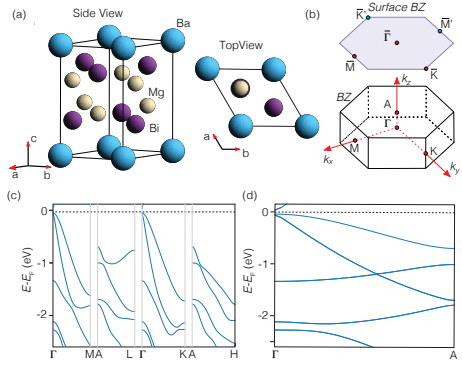}
		\caption{\label{fig:Fig1}(a) Side and top view of the structure and unit cell of BaMg$_2$Bi$_2$ crystal, showing the typical CaAl$_2$Si$_2$ structure. (b) Resulting Brillouin zone with high symmetry directions indicated. The surface Brillouin zone is also indicated. (c) Bulk electronic structure along the in-plane high symmetry directions. (d) Same as (c) but along the out of plane direction.}
	\end{figure}

The interplay between crystalline symmetry and band inversion -- often facilitated by spin-orbit coupling -- plays a central role in the emergence of novel topological phases of matter \cite{Wojec_2014, Narang_2021, Mazzola_2023, Mazzola_2023b}. Among these, topological Dirac semimetals (TDSs) represent a distinct class, differing fundamentally from conventional topological insulators. TDSs materials have garnered significant interest due to their exceptionally high carrier mobility \cite{Neupane_2014, Fujioka_2019}, the presence of surface Fermi arcs that give rise to unconventional transport phenomena \cite{Wieder_2020, Potter_2014, Yang_2015}, and their typically large, non-saturating magnetoresistance \cite{Leahy_2018, He_2021}. From a band structure perspective, a TDS is characterized by massless Dirac fermions, manifested as linear band crossings (Dirac nodes) at or near the Fermi level. Importantly, under external perturbations such as strain or applied magnetic fields, topologically trivial Dirac nodes can split into Weyl nodes with opposite non-trivial chiralities, thereby realizing a topological transitions. These tunable responses make TDSs highly promising for applications in next-generation electronic and spintronic devices.

         \begin{figure*}
		\includegraphics[width=\textwidth]{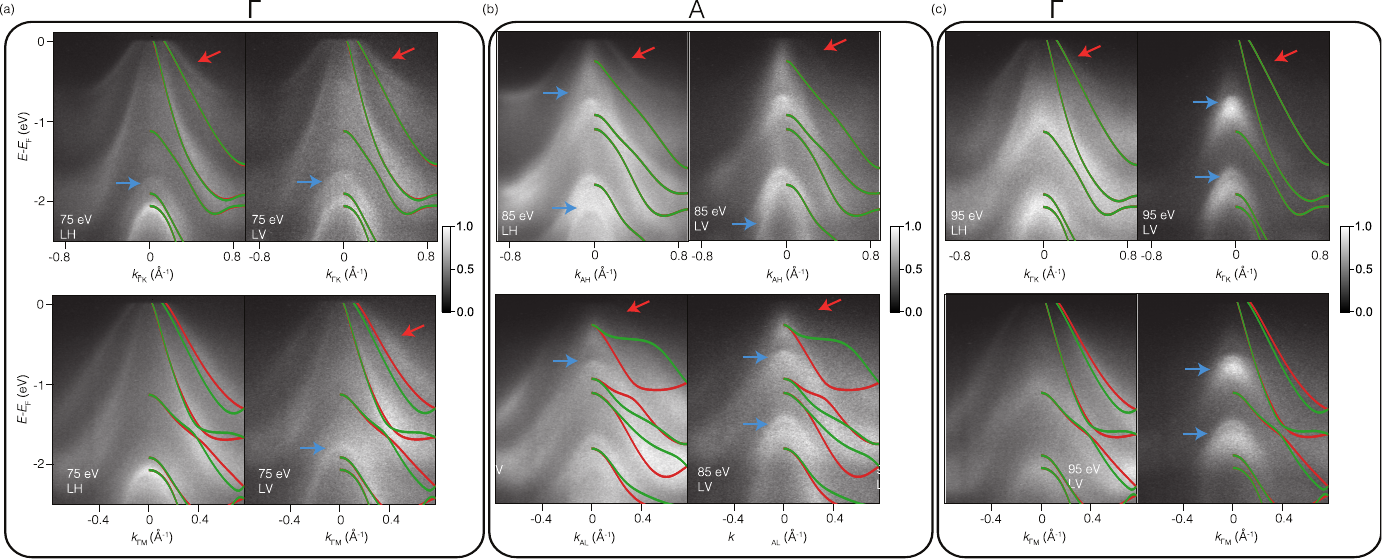}
		\caption{\label{fig:Fig2}(a) $E$--$k$ dispersions of BaMg$_2$Bi$_2$ along the $\Gamma$--$K$ (top) and $\Gamma$--$M$ (bottom) directions for both LH and LV polarizations. Spectra were acquired with 75~eV photons, corresponding to a bulk $\Gamma$ point. (b) Same dispersions acquired with 85~eV photons, corresponding to the $A$ point. (c) Spectra obtained with 95~eV photons, corresponding to the next bulk $\Gamma$ point. Red and blue arrows highlight additional features not previously reported.}
	\end{figure*}

BaMg$_2$Bi$_2$ crystallizes in the CaAl$_2$Si$_2$-type structure, corresponding to space group $P\overline{3}m1$ (No.~164). This crystallographic configuration enforces a $C_3$ rotational symmetry, which plays a pivotal role in stabilizing the system's only Dirac cone --located at the center of the Brillouin zone and linearly dispersing across the Fermi level (See Fig.~\ref{fig:Fig1}) \cite{Liu_2022}. Crucially, despite hosting a symmetry-protected Dirac node, BaMg$_2$Bi$_2$ is topologically trivial, thus providing a clean model platform for investigating textbook Dirac fermions in the absence of additional topological features \cite{Takane_2021}. Moreover, this structure exhibits considerable chemical flexibility; for instance, substituting Ba with other alkaline earth or rare-earth elements \cite{Kundu_2022, Kundu_2022_bis, Zhang_2022} can effectively modulate the spin-orbit coupling strength by altering the local atomic environment. Such tunability is particularly valuable for exploring topological phase transitions in this material family.

Previous experimental studies have explored the electronic structure of BaMg$_2$Bi$_2$ using angle-resolved photoemission spectroscopy (ARPES), confirming the Dirac nature of this compound through the observation of dispersing massless fermions, and directly probing the Dirac point via K-deposition \cite{Takane_2021}. However, measurements conducted at relatively high photon energies suffer from limited resolution, which hinders a clear visualization of other electronic states. In particular, due to the increased bulk sensitivity at high photon energies, it is challenging to detect surface states. Moreover, electronic doping via alkali deposition modifies the surface potential and suppresses intrinsic surface states. Here, by combining high-resolution angle-resolved photoemission spectroscopy (ARPES) with variable light polarization and photon energy \cite{Bigi_2025, Bigi_nematic}, together with density functional theory (DFT) calculations, we complete the picture of the electronic structure. Our study reveals the presence of additional states—some arising from $k_z$ broadening artifacts, and others which we attribute to the surface state manifold. Our polarization-dependent measurements in the low-energy regime shed light on additional features of the electronic structure which can be important/relevant for transport in BaMg$_2$Bi$_2$.

High-quality single crystals of BaMg$_2$Bi$_2$ were grown via a self-flux method using the constituent elements in a 1.5:5:9 molar ratio (Ba:Mg:Bi). The mixture was loaded into an alumina crucible, sealed in an evacuated quartz tube, and heated to 900°C for 12 h. It was then slowly cooled to 650°C at a rate of 1°C/h. The crystals were separated from the flux by centrifugation. The resulting crystals exhibited smooth surfaces and a pronounced metallic lustre. ARPES measurements were carried out at the CASSIOPEE beamline of the synchrotron radiation facility SOLEIL (Paris), using both linear vertical (LV, \textit{s}-polarized) and linear horizontal (LH, \textit{p}-polarized) polarizations. In the geometry of this experimental setup, LV-polarized photons have their electric field vector entirely confined within the sample plane, while LH polarization introduces both in-plane and out-of-plane components of equal magnitude. The resolution of the beamline is better than \SI{10}{\milli\eV}. ARPES was performed under ultrahigh vacuum conditions (UHV) at \SI{20}{\K} base temperature. The samples were prepared in a glovebox environment by affixing them to the sample plate using two-component silver epoxy. They were subsequently top-posted and cleaved under UHV conditions at \SI{20}{\K}. This procedure ensures that the surface properties are not destroyed by low vacuum and adsorption of impurities, a situation that can often occur if samples are cleaved at higher temperature and then cooled down.

\begin{figure*}
		\includegraphics[width=0.6\textwidth]{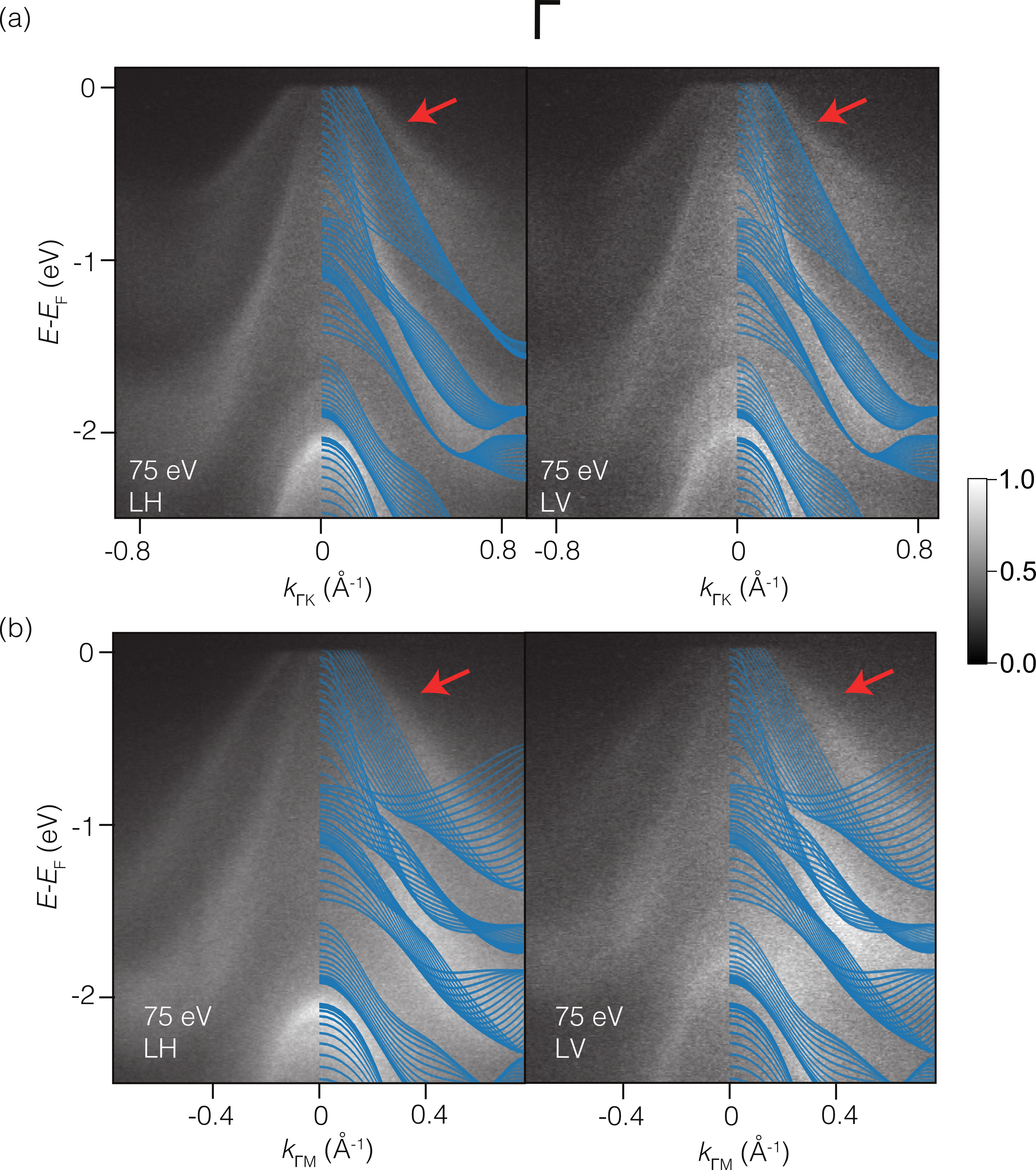}
		\caption{\label{fig:Fig3} (a) Electronic structure of BaMg$_2$Bi$_2$ along the $\Gamma$-K and (b) $\Gamma$-M directions for both light polarizations indicated in the figure. DFT calculations have been overlaid and the gray color indicates the projected bulk continuum.}
	\end{figure*}

In addition, first-principles calculations were performed using the QuantumESPRESSO DFT package. The self-consistent calculations were carried out on a $6\times6\times6$ k-point grid with a 60 Ry energy cutoff, employing fully relativistic projector augmented wave (PAW) pseudopotentials. It is important to note that, due to the lack of six-fold rotational symmetry in the crystal, there are two inequivalent M-points and two inequivalent K-points, which are related by a 60° rotation of M and K around the (001)-axis. For the (001)-slab model, where the (001)-surface is truncated by vacuum, the system retains only one spatial symmetry, as the truncation breaks inversion symmetry. The remaining symmetry operation is a 180° rotation about the (110)-axis combined with inversion. This breaking of inversion symmetry could permit the existence of spin-split surface states. Time-reversal symmetry is preserved in both the bulk and slab geometries.

Leading from our combined experimental and theoretical data, in Fig.~\ref{fig:Fig2} we present the bulk‐calculated electronic structure alongside ARPES spectra acquired at variable photon energies and polarizations. A change in photon energy results in a variation of the probed $k_z$ (going with the square root of the photon energy), thus allowing to probe different areas of the 3D Brillouin zone. Overall, and consistently with  Ref. \cite{Takane_2021}, there is general agreement between the ARPES data and the DFT calculations: we reveal the main features of the valence band in BaMg$_2$Bi$_2$. The outermost and rapidly dispersing band crossing $E_F$ and forming the Dirac point in the conduction band, as shown by Takane \textit{et al.} \cite{Takane_2021}, is often saddled by unfavourable matrix elements, and is therefore partially visible only at some photon energies. The overlap between experimental and calculated bands displays some discrepancies in terms of energy position of the main features, but overall the calculations follow the experimentally revealed dispersion. We also note the relevant broadening of most of the measured electronic states. This may be attributed to intrinsic disorder effects at the cleaved surface of BaMg$_2$Bi$_2$: surface relaxation and vacancy formation are likely to alter the on-site energies and the degree of electronic correlation at the surface, resulting in energy shifts and spectral broadening observed experimentally. The pronounced surface sensitivity in the \SIrange{75}{95}{\eV} photon energy range makes our measurements relatively prone to revealing such effects. Nevertheless, as indicated by the red and blue arrows in Fig.~\ref{fig:Fig2}, the major and striking difference is that several bands at each $k_z$ value are neither captured by DFT nor reported in earlier experiments. The measurements span a photon‐energy range that covers multiple Brillouin zones; for brevity, we show one repetition in the main text, while additional energies are provided in the Supplementary Information (Supplementary Figs.~1 and 2). Because the unidentified features persist across the entire $k_z$ range and for both light polarizations, they cannot be attributed to photoemission matrix‐element effects; instead, they point to an intrinsic characteristic of the electronic structure of BaMg$_2$Bi$_2$. These features are evident along both the $\Gamma$–K and $\Gamma$–M high‐symmetry directions.

\begin{figure*}
		\includegraphics[width=0.5\textwidth]{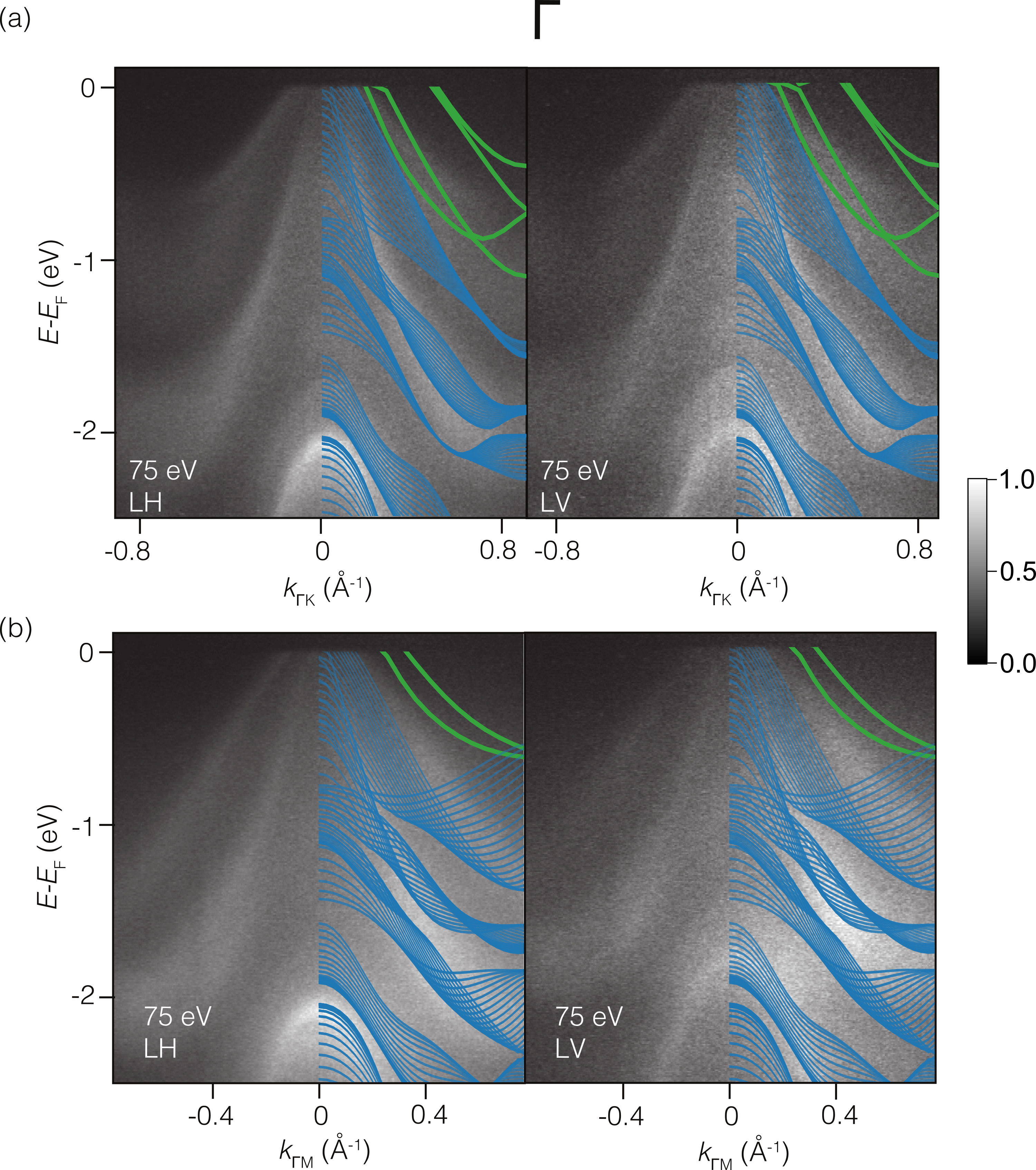}
		\caption{\label{fig:Fig4} (a) Electronic structure of BaMg$_2$Bi$_2$ along the $\Gamma$-K and (b) $\Gamma$-M (bottom) directions for both light polarizations indicated in the figure. DFT calculations have been overlaid and the gray colour indicates the projected bulk continuum. The green colour indicate the surface states, calculated as described in the text.}
	\end{figure*}

Mapped in Fig.~\ref{fig:Fig2}, the calculated dispersions along $\Gamma$–-K directions, as well as the $\Gamma$–-M and the symmetry‐inequivalent $\Gamma$--M$'$ directions (due to the intrinsic $C_3$ symmetry of BaMg$_2$Bi$_2$) are shown in red and green, respectively. Given the photon beam’s lateral size of approximately \SI{100}{\micro\metre}, domains rotated by \ang{180} are likely probed simultaneously. The resulting superposition of $\Gamma$--M and $\Gamma$--M$'$ contributions renders the experimental spectra more symmetric and partially broadens the linewidths, although differences -- especially in the intensity of some bands -- are clearly visible between the two symmetry-inequivalent spectra. To account for this, we overlay both inequivalent cuts in the theoretical plot. However, domain averaging alone cannot reproduce the additional spectral weight highlighted above, indicating that further mechanisms must be considered.

Examining the data presented in Fig.~\ref{fig:Fig2}, we note that the photon energy spans a full $k_z$ period, covering the path from $\Gamma$ to the next $\Gamma$ point. The bands reach their minimum binding energy at the center of the Brillouin zone and their maximum at the A point. Therefore, the features marked by the red arrows in Fig.~\ref{fig:Fig2} are unexpected, as they appear above the calculated bands in Fig.~\ref{fig:Fig2}a and ~\ref{fig:Fig2}c. In contrast, additional features appearing at lower binding energies -- still previously unobserved -- could be more straightforwardly explained. We now focus on these latter bands, highlighted by the blue arrows.

Remarkably, despite an overall modulation and redistribution of intensity, all bands remain mostly visible across the entire photon energy range. Although this may initially seem inconsistent with previous studies, the observed phenomenology can be reconciled by considering the strong $k_z$ broadening in both the ARPES spectra and the bulk-projected simulations. Experimentally, $k_z$ broadening manifests as a general filling of spectral intensity over the entire $k_z$-dependent dispersion range. This effect arises because, despite tuning to a specific photon energy, multiple $k_z$ values are effectively probed simultaneously. Such a condition is common in materials with strongly three-dimensional electronic structures and large c-axis lattice constant \cite{Pakpour_2020, DeVita_2025}, and it is therefore not surprising that it occurs in BaMg$_2$Bi$_2$ \cite{Deller1977}.

Driven by this interpretation, our bulk-projected DFT calculations successfully reproduce the spectral broadening observed in the ARPES data. This comparison is shown in Fig.~\ref{fig:Fig3} for the bulk $\Gamma$ point, along both the $\Gamma$--K and $\Gamma$--M directions. As clearly illustrated, the filling of the electronic states seen in the ARPES measurements finds a simple yet compelling explanation: several of the states that appeared experimentally -- previously unreported -- are now accounted for. However, the features highlighted in red remain unexplained within the framework of bulk-projected DFT, suggesting a non-bulk origin.

As a means of verification, we performed surface-state calculations using a slab geometry that is infinite in the \textit{x}- and \textit{y}-directions but truncated along \textit{z} (the (001)-axis). The slab consisted of 24 atomic layers with a \SI{12}{\angstrom} vacuum separating the two surfaces. In Fig.~\ref{fig:Fig4}, the projected bulk continuum is shown in light blue (obtained by projecting the bulk band structure onto the (001)-plane), while the surface states extracted from the slab calculation are highlighted in green (identified based on their probability density). We confirm the localization at the surface of these bands by projecting their probability density onto the BaMg$_2$Bi$_2$ slab; one example for the lowest-energy surface state is shown in Supplementary Information, Fig.~S3. Polarization-dependent ARPES spectra show that these states are visible under both LH and LV light, implying contributions from orbitals with both in-plane and out-of-plane symmetry. This behavior is consistent with hybridization between Bi-p$_z$ and Bi-p$_(x,y)$ orbitals, slightly modified by the broken inversion symmetry at the surface.

Since the $\Gamma$ and M points are time-reversal invariant momenta, the surface states remain degenerate at these points, meaning that their spin splitting vanishes. However, this does not hold at K, as time-reversal symmetry alone does not enforce degeneracy there due to the crystal symmetry. The presence of spin splitting, as shown \textit{e.g.} for the \textit{x} and \textit{y} components of the spin (the \textit{z} component is zero, enforced by crystal symmetry) in Fig.~S4, is therefore not unexpected. Examining the number of Fermi-level crossings of the surface states between high-symmetry points, we find that it is always even. This indicates that the surface states are topologically trivial \cite{PhysRevB.76.045302}. If they were nontrivial, we would expect them to exhibit an odd number of Fermi-level crossings and ultimately connect the bulk conduction and valence bands, whereas in the case of BaMg$_2$Bi$_2$ they touch at the $\Gamma$ point.

The observed surface states arise from the intrinsic potential discontinuity at the interface between the crystal and the vacuum. The cleavage along the (001) plane breaks inversion symmetry and modifies the local coordination of the surface Bi atoms, leading to a redistribution of charge and a rehybridisation of Bi-\textit{p} and Mg-\textit{s} orbitals. This creates localized, topologically trivial states confined to the outermost layers of the slab. Their existence is therefore a natural consequence of the surface termination rather than of a bulk topological mechanism manifesting at the surface, yet they represent a key ingredient in fully understanding the electronic properties of BaMg$_2$Bi$_2$.

In summary, we have conducted a comprehensive investigation of the electronic structure of BaMg$_2$Bi$_2$ using high-resolution ARPES, combined with polarization- and photon-energy-dependent measurements, supported by both bulk and slab-based DFT calculations. Our results confirm that BaMg$_2$Bi$_2$ hosts a symmetry-protected Dirac node at the $\Gamma$ point, stabilized by the underlying threefold rotational symmetry of the CaAl$_2$Si$_2$-type structure. While the Dirac dispersion has been previously reported, our experiments reveal additional features in the spectral function—some attributable to strong out-of-plane momentum ($k_z$) broadening and others not accounted for by calculations of the bulk electronic structure.

Through bulk-projected simulations, we demonstrate that the observed intensity filling in the photoemission spectra is a natural consequence of momentum broadening along the out-of-plane direction in a material with significant three-dimensional electronic character. However, distinct features that remain unexplained within the bulk framework are successfully captured by calculations performed on a finite slab geometry. These slab-derived bands, while topologically trivial, constitute an essential part of the low-energy electronic structure of BaMg$_2$Bi$_2$. Possibly, surface scattering and magnetoresistance may be influenced by the presence of surface states. Moreover, Liu et al. \cite{Liu_2022} have revealed 2D superconductivity below $\sim\SI{5}{\K}$ in this material. The presence of surface states results in a depth-dependent density of states near $E_F$, which is likely to contribute at least partly to the stabilization of the 2D superconducting state. Subtle structural modifications at the surface of BaMg$_2$Bi$_2$, such as atomic substitution or hybridization engineering of the outermost Bi atoms, may thus represent a promising pathway to tune the superconducting state via surface states manipulation. 

Altogether, our work not only reinforces BaMg$_2$Bi$_2$ as a model Dirac semimetal with topologically trivial character, but also uncovers additional electronic states -- originating from both bulk and surface effects -- that have gone undetected in previous studies. These findings highlight the importance of combining energy-, momentum-, and polarization-resolved spectroscopy with carefully constructed theoretical models in order to fully capture the complexity of the electronic landscape in candidate Dirac materials.

\textbf{Acknowledgments:}

F.M. greatly acknowledges the NFFA-DI funded by the European Union – NextGenerationEU, M4C2, within the PNRR project NFFA-DI, CUP B53C22004310006, IR0000015. A.D.V., R.E., and T.P. acknowledge the Deutsche Forschungsgemeinschaft (DFG) within Transregio TRR 227 Ultrafast Spin Dynamics project B07, the Max Planck Society, and BERLIN QUANTUM, an initiative endowed by the Innovation Promotion Fund of the city of Berlin. Work at Gdansk Tech. was supported by Platinum Establishing Top-Class Research Teams project (DEC-2/2/2023/IDUB/I.1B/Pt). JAM and CVBN  gratefully acknowledge support from DanScatt (7129-00018B). 

 \bibliography{bibliography}
	
\end{document}